\documentclass[12pt]{article}
\usepackage{epsfig}
\usepackage{psfig}
\usepackage{amssymb}
\usepackage{scicite}

\newenvironment{sciabstract}{%
\begin{quote} \bf}
{\end{quote}}

\newcounter{lastnote}

\title{Iron Emission Lines from Extended X-ray Jets in SS~433: Reheating of
Atomic Nuclei}

\author
{Simone Migliari,$^{1\ast}$ Rob Fender,$^{1\ast}$ Mariano M\'endez,$^{2}$\\
\\
\normalsize{$^{1}$Astronomical Institute `Anton Pannekoek', University of
Amsterdam, Kruislaan 403,}\\
\normalsize{1098 SJ Amsterdam, The Netherlands}\\
\normalsize{$^{2}$SRON, National Institute for Space Research, 3584 CA
Utrecht, The Netherlands}\\
\\
\normalsize{$^\ast$To whom correspondence should be addressed; E-mail:} \\
\normalsize{migliari@astro.uva.nl (S.M.); rpf@astro.uva.nl (R.F.)}
}

\date{}

\begin{document}

\maketitle 
\baselineskip24pt

\begin{sciabstract}
Powerful relativistic jets are among the most ubiquitous and
energetic observational consequences of accretion around
supermassive black holes in active galactic nuclei and neutron
stars and stellar-mass black holes in x-ray binary (XRB) systems. 
But despite more than three decades of study, the structure and composition of
these jets remain unknown. 
Here we present spatially resolved
x-ray spectroscopy of arc second-scale x-ray jets from the XRB SS
433 analized with the Chandra advanced charge-coupled device imaging
spectrometer. These observations reveal evidence
for a hot continuum and Doppler-shifted iron emission
lines from spatially resolved regions. Apparently, in situ reheating of
the baryonic component of the jets takes place in a flow that moves with
relativistic bulk velocity even more than 100 days after launch from the
binary core.  
\end{sciabstract}

\noindent 

\begin{figure}[!P]
     \psfig{figure=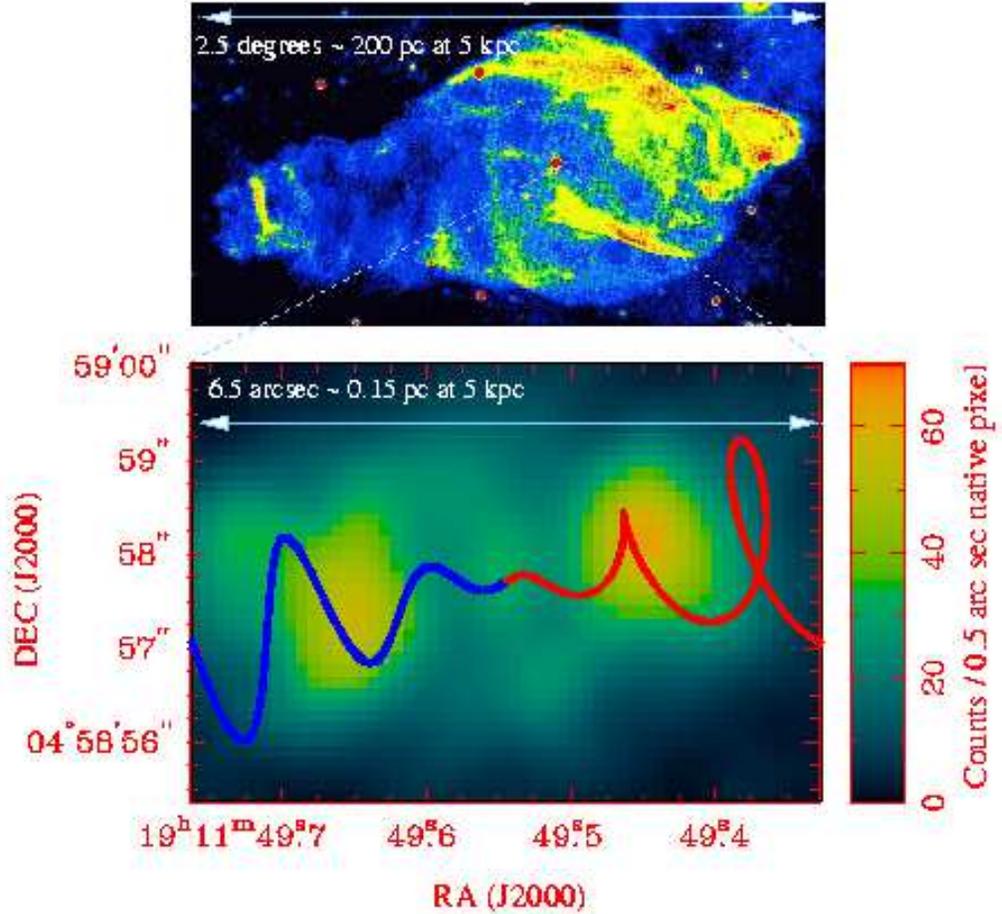,angle=-90,width=18cm} \caption{{\bf (Bottom)}
    Image of SS 433 with Chandra ACIS-S, rebinned to one-quarter
    native pixel size and 
    smoothed with a 0.5-arc sec Gaussian (comparable to the resolution
    of the telescope). The predicted projection on the
    sky of the jet precession cycle is indicated. The blue line refers to the jet that, for
    the most of the time, is approaching Earth (it is receding only $\sim 16$
    days over the 162-day
    precession cycle), and the red line refers to the (mostly) receding
    jet. The x-ray emission lies in the
    jet path and, while concentrated in two lobes, shows weaker
    emission extended beyond these to larger distances at the
    predicted position angles. {\bf (Top)} Comparison with the surrounding W50 radio
    nebula ({\it 9}) shows the correspondence in alignment of the
    structures, both of which reveal the signature of the jet
    precession ({\it 19}). DEC, declination; RA, right ascension.  
      }
    \label{fig1}
\end{figure}

The jets of XRB SS 433 are well established, quasi-persistent and the best
example of regular precession of the jet axis ({\it 1--3}).  Extensive
observing campaigns have placed constraints on the emission physics, geometry and scales
associated with the baryonic component of the jets of SS 433.  The generally
accepted model for the thermal component of the jet is an adiabatic cooling
model ({\it 4, 5}).  This model assumes that the emitting matter is moving
along ballistic trajectories and expanding adiabatically in a conical outflow,
and that the temperature and matter density are only a function of the
distance from the core. The matter is ejected (the physical processes of
ejection are still uncertain) with a temperature of $>10^{8}$ K
(mean thermal energy per particle kT$>10$ keV) ({\it 4--6}), comparable to
that measured for the inner regions of accretion disks. 
Close to the binary core, at distances less than $\sim 10^{11}$~cm, 
this jet plasma emits high-temperature continuum and Doppler-shifted lines
in the soft x-ray band ({\it 6}).  The jet subsequently emits Doppler-shifted
optical emission lines that are constrained to originate at distances of
less than $\sim 10^{15}$ cm because of a lack of measurable offsets between blue- and
red-shifted beams. Neither line-emitting region has been spatially resolved
from the core of SS 433. Furthermore, the temperature of the optical 
line-emitting region is limited to less than $\sim 10^{4}$ K by the absence of higher
excitation lines such as He {\small II} and C {\small III}/N {\small III}
({\it 7}).  In the adiabatic model the temperature decreases with the radial
distance $R$ from the core as temperature T$\propto$R$^{-4/3}$; thus, at distances greater
than $\sim 10^{16}$ cm, the matter is expected to be at a temperature of
$<100$ K, too cool to thermally emit x-ray and optical radiation. Besides
this thermal radiation, there is nonthermal synchrotron radiation, which
traces the distribution of relativistic electrons and the magnetic field in the
jet, observed in the radio band at distances between $10^{15}$ and $10^{17}$ cm
({\it 2}). Further out, the jet is not detected again until degree-scale (i.e.,
at distances of $\sim 10^{19}$ to $10^{20}$ cm from the binary core) termination
shocks are observed, deforming the surrounding W50 nebula (Fig.~\ref{fig1}, top), which
is thought to be a supernova remnant associated with the formation of
the neutron star or black hole in SS 433 ({\it 8, 9}).

Marshall {\em et al.} ({\it 6}) have recently analyzed the x-ray line emission from
SS 433 with the Chandra high energy transmission grating spectrometer
(HETGS). They refined the adiabatic cooling model, limiting the thermal x-ray
emission region to within $\sim 10^{11}$ cm of the jet base. They also
discovered faint arc sec-scale x-ray emission on either side of the core, but
they were not able to measure any emission lines and concluded that the jet
gas would be too cool at such a distance to emit thermal x-rays.

We have analysed a 9.6-ks observation from 27 June 2000 with the Chandra
advanced charge-coupled device (CCD) imaging spectrometer (ACIS)-S. The reduced image
(Fig.~\ref{fig1}) resolves arc sec-scale x-ray jets approximately
symmetrically placed on either side of a weaker core and aligned with both the
sub-arc sec radio jets and degree-scale structures in W50. Analysis of the
image indicates that photon pile-up ({\it 10}) in the core artificially
reduced the measured count rate there, although according to
the orbital ephemeris of Dolan {\em et al.} ({\it 11}), our observation was close to
x-ray eclipse, phase $\sim 0.9$ ({\it 12}), which may also have reduced the
core flux.  At a distance of 5 kpc, these x-ray jets correspond to physical
scales of $\geq 10^{17}$ cm and soft x-ray (2 to 10 keV) luminosities of 3$\times 10^{33}$ to
4$\times 10^{33}$ erg s$^{-1}$ each (about 3\% of the average source total
luminosity).

\begin{figure}[!P]
      \psfig{figure=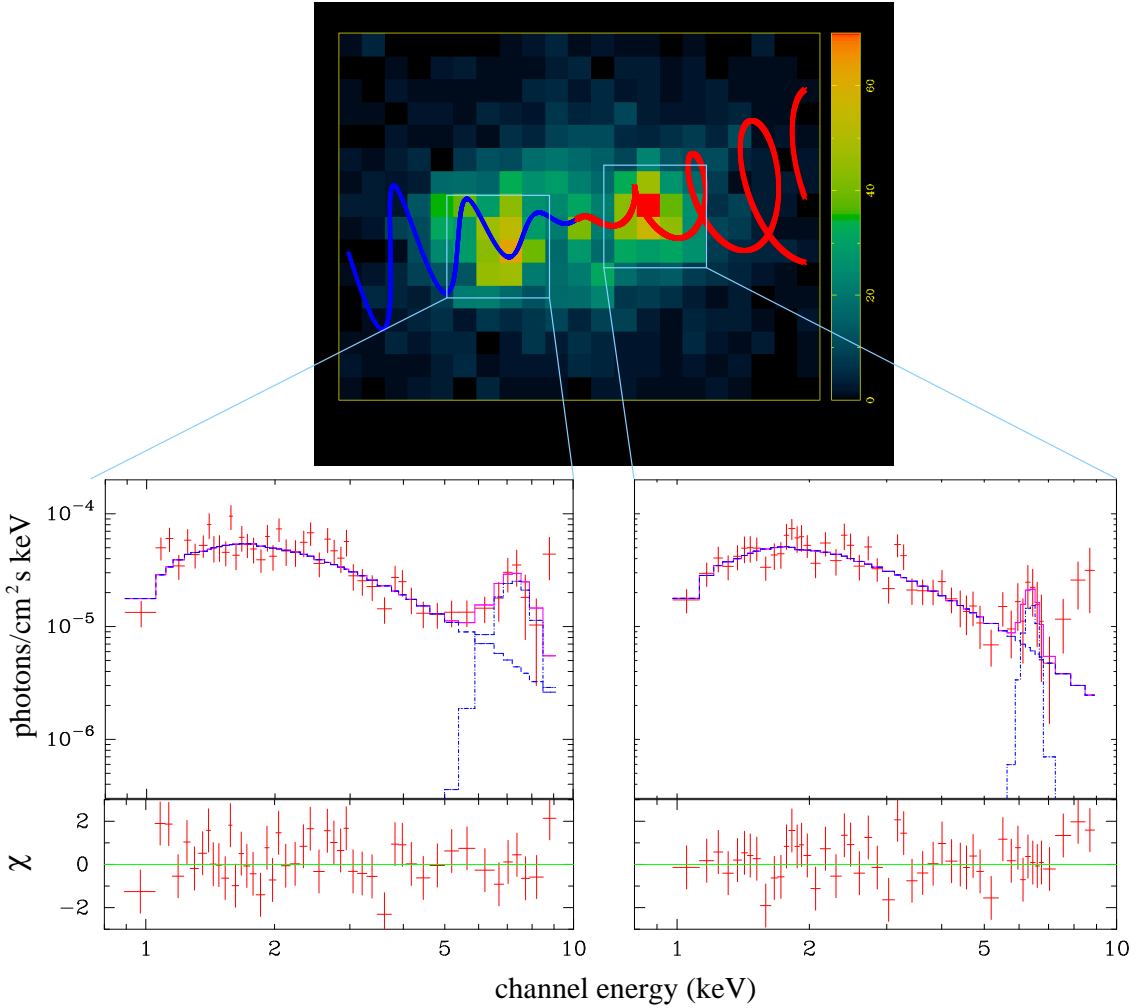,width=15.0cm,angle=0}
    \caption{{\bf (Top)} Raw (i.e., not regridded) Chandra ACIS-S image of SS
433 with the projected precession cycle of the jets superimposed on it,
indicating the regions of the jet we isolated and analyzed. {\bf (Bottom)} The
two spectra of the lobes (crosses), east (left) and west (right), with the
fitted model [absorbed bremsstrahlung (dashed line) and Gaussian emission line
(dashed-dotted line)] and residuals ($\chi$). The error bars represent the
1$\sigma$ confidence interval for each measurement; the residuals are
plotted in units of $\sigma$.}
    \label{fig2}
\end{figure}

We were able to spatially isolate and take x-ray
spectra of the two jets.  The spectra
of the two lobes between $\sim 0.8$ and 10 keV (Fig.~\ref{fig2}) can be fit 
by a bremsstrahlung continuum with a temperature of $\sim 5$ keV and
Gaussian emission lines at $\sim 7.3$ keV for the approaching (east) jet and at
$\sim 6.4$ keV for the receding (west) jet (Table 1). In both
spectra there is an indication of additional emission, possibly also
lines, at higher energies, but this is near to the limit of the
Chandra sensitivity, and given our relatively low count rates, we did
not attempt to fit this component. A hydrogen column density of $\sim
10^{22}$ cm$^{-2}$ was found for both lobes, consistent with the column density
found by Marshall {\em et al}. ({\it 6}) 
for the core of SS~433. A power
law with a photon index of $2.1\pm 0.2$ fits the continuum equally
well, and on the basis of our current data, we cannot distinguish between a
power law and bremsstrahlung for the continuum emission.

\begin{figure}
      \psfig{figure=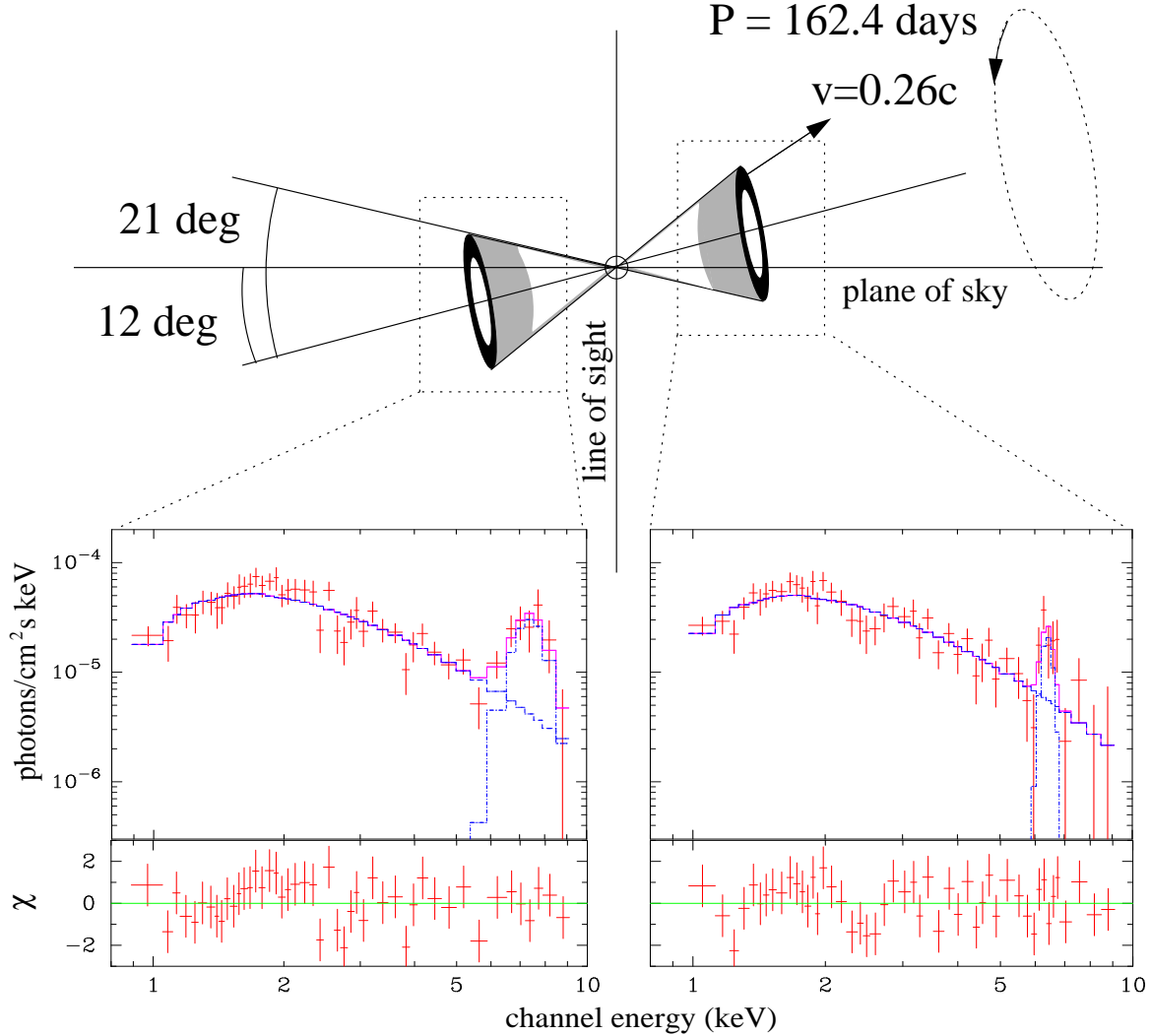,width=14.0cm,angle=-90}
    \caption{Simulated spectra (crosses), 
    integrated over one entire precession cycle and fit for the same
    integration time, with the same
    binning and model [absorbed bremsstrahlung (dashed line) and Gaussian
    emission line (dashed-dotted line)] as for our observation in
    Fig. 2. 
    Residuals ($\chi$) of the fit are also plotted. The error bars represent
    the 1$\sigma$ confidence interval for each measurement; the residuals are
    plotted in units of $\sigma$. Parameters   
    for the jet precession model (jet velocity $v=0.26c$, opening angle
    $\theta = 21^{\circ}$, precession axis angle $i = 78^{\circ}$, 
    period $P = 162.4$ days) are from ({\it 3}). 
    For each of ten precession phases, the observed
    line energy $E_{\rm obs} = \delta E_{\rm rest}$ and
    Doppler-boosted peak intensity $I_{\rm obs} = \delta^{2} I_{\rm
    rest}$ were calculated and summed to produce the resultant
    line profiles (the model can be visualized as simulating the
    integrated line emission of the shaded region in the schematic).
    The normalization of the summed lines was chosen to be the
    same as that in our observations, and the rest energy of the line
    was chosen to correspond to that of Fe XXV k$\beta$  (i.e., 7.06 keV).}
    \label{fig3}
\end{figure}

The significantly different line centroid energies measured for the east and
west jets suggest that they may originate in the moving jet material.  To test
this hypothesis, we developed a model that sums the emission from ten
equally spaced phase bins ({\it 13}) around the precession cycle, calculating
for each bin the Doppler shift $\delta$ ({\it 14}) and Doppler boosting of the
line peak intensity $\delta^{2}$ (assuming that the jets are intrinsically symmetric
about the precession axis and east to west). A simulation based on this model
for a 10-ks observation, assuming a rest energy for the line of 7.06 keV,
corresponding to Fe XXV k$\beta$ and using the same normalizations for the
Gaussians as found in our observation, corresponds well to our data
(Fig.~\ref{fig3} and Table 1).

The line energies and widths we obtain in our simulation are a
good match to those measured in our observation (Fig.~\ref{fig2}
and~\ref{fig3}). Thus, we infer that the jet is emitting around the
precession cone at a large distance from the core. Our
model predicts twin-peaked lines, due to the distribution of
Doppler shifts as a function of time, with a stronger blue peak (due
to Doppler boosting); however, this twin-peak is not seen in the 10-ks
simulation, which have a relatively low signal-to-noise ratio.
However, if our model is correct, then additional observations should
reveal this structure. Although we cannot use
our model to conclusively establish that the Doppler-shifted line is
Fe XXV k$\beta$, we do need a transition of hot, highly ionized Fe
in order to achieve a high enough energy for the line in the east jet.
We cannot exclude an origin for the $\sim
6.4$ keV line in the west jet from neutral Fe; however, we cannot
envisage a plausible geometry whereby we are observing neutral Fe at
6.4 keV in the west jet and the same line extremely blue-shifted in
the east jet. A further prediction of this model is that the east jet
should be stronger than the west jet as a result of the overall larger
Doppler boosting. The ratio of 2 to 10 keV fluxes (line and continuum) from the
east to west jets is $\sim 1.3$, approximately what would be expected for the
model considered here. 
Although we cannot explicitly
show that the source of the continuum emission arises in the moving
jet gas, the measured bremsstrahlung temperature is consistent with an
origin in the same plasma as the high-excitation Fe lines.  We
therefore conclude that we observed lines, and probably also
continuum, from a hot ($>10^{7}$K) plasma which is moving with a
similarly high velocity to that measured for the ``inner'' jets ({\it 6}),  
$\sim 0.26c$ (where $c\sim 3\times 10^{10}$ cm s$^{-1}$ is the speed of
light), on scales of $10^{17}$ cm from the binary core. 

This is contrary to the adiabatic cooling model ({\it 4--6}) 
which at that distance predicts that the baryonic
component will have cooled to the ambient interstellar medium temperature;
therefore, the jet has to re-heat itself. The most plausible energy source is
the kinetic energy ($\sim 10^{38}$ ergs/s) associated with the bulk motion of
the jet ({\it 6}).  
Because we
observed optical emission lines coming from matter with temperatures of $<
10^{4}$ K at a distance of $\sim 10^{15}$ cm, this re-heating must
occur between $\sim 10^{15}$ and $10^{17}$ cm from the core.

Alternatives to this scenario can be considered. A power law fits
the continuum well, and by analogy with the extended x-ray jets seen
in active galactic nuclei (AGN) ({\it 15}), 
we cannot exclude an inverse-Compton or synchrotron
origin for the continuum component. However, because both of these emission
mechanisms require a population of hot electrons, they also imply
reheating at those large distances from the central source.
Regardless of which form of reheating is associated with the continuum
spectrum, the emission lines require a baryonic component. Reflection of the
core emission does not seem possible because
the line flux we observed at $10^{17}$ cm is only a factor of $\sim 10$
less than that observed to originate at $10^{12}$ cm ({\it 6}),
and yet our images indicate it does not subtend a significantly larger solid
angle than the jets on smaller scales. Furthermore, the line to continuum
flux ratio we observed is an order of magnitude greater than in the
core [$F_{\rm line} / F_{\rm continuum}$ is $\sim 1$ for our east jet, but
is $\sim 0.1$ for the 7.3-keV line in Chandra HETGS observations
({\it 6})]. 
The lines from these extended jets are still an
order of magnitude weaker than the moving core x-ray lines
({\it 5, 6}), 
and would not be expected to show
measurable changes in their Doppler shifts (unless, e.g., a major
outburst from SS 433 resulted in enhanced emission at some particular
precession phase). Nevertheless, they should be present as weak
features in spectra that do not spatially resolve the core and jets.

Reheating (in situ acceleration) of a leptonic (electron and possibly
also positron) 
component of jet plasma, presumably by conversion of the jet kinetic
energy, is commonly invoked to explain synchrotron or inverse Compton
energy emission on large scales in the jets of AGN ({\it 15-17}). 
The reheating of the baryonic plasma
that we observed in SS 433 requires a similar tapping of the
bulk kinetic energy of the flow, by processes that are able
to act on atomic nuclei. In the internal shock model for blazars
({\it 16}), 
a slightly varying jet speed results in shocks
downstream in the jet flow. In the context of this model, random discrepancies
of about 15\% in the velocity of the jets from SS 433, inferred from long-term optical
monitoring ({\it 3}),  
could produce shocks a few hundred days
downstream (for a precession period of 162 days and a mean jet
velocity of $0.26c$), comparable to what we observed ({\it 18}) .  

To conclude, these observations spatially resolve a line-emitting
region in a relativistic jet. They reveal that
the hot x-ray emitting region is still moving relativistically on
physical scales that are orders of magnitude larger than previously inferred,
with no evidence for substantial deceleration, supporting models for
extended x-ray jets in AGN that require bulk relativistic motion
on large physical scales ({\it 15}). Furthermore, this observation
demonstrates that particle reacceleration in a
relativistic jet, previously inferred only to act on the leptonic
component, can act also on atomic nuclei. Comparison of this highly
super-Eddington, mass-loaded jet with those of other x-ray binary systems and AGN can
provide unique insights into the matter and energy content of
relativistic jets and hence the coupling of accretion and outflow in
conditions of extreme gravity, pressure, and energy density.

\begin{quote}
{\bf References and Notes}

\begin{enumerate}

\item G. O. Abell and B. Margon, {\it Nature} {\bf 279}, 701 (1979).  

\item  R. M. Hjellming and K. J. Johnston, {\it Nature} {\bf 290},
100 (1981).  

\item S. S. Eikenberry, {\it et al.}, {\it Astrophys. J.} {\bf 561},
1027 (2001).  

\item W. Brinkmann, N. Kawai, M. Matsuoka, H. H. Fink, {\it
Astron. Astrophys.} {\bf 241}, 112 (1991).   

\item T. Kotani, N. Kawai, M. Matsuoka, W. Brinkmann,
{\it Publ. Astron. Soc. Pac.} {\bf 48}, 619 (1996).   

\item H. L. Marshall, C. R. Canizares, N. S. Schulz, {\it
Astrophys. J.} {\bf 564}, 941 (2002).      

\item J. Shaham, {\it Vistas Astron.} {\bf 25}, 217 (1981).

\item W. Brinkmann, B. Aschenbach, N. Kawai, {\it
Astron. Astrophys.} {\bf 312}, 306 (1996).

\item G. M. Dubner, M. Holdaway, W. M. Goss, I. F. Mirabel, {\it
Astron. J.} {\bf 116}, 1842 (1998).  

\item Pile-up is an effect by which two or more photons that hit the same
pixel on the CCD in one integration are counted as a single photon. If the
summed energy is greater than the ACIS threhold ($\sim 15$ keV), the detection
is rejected, and no counts are recorded. 

\item J. F. Dolan, {\it et al.}, {\it Astrophys. J.} {\bf 327}, 648 (1997).

\item The orbital phase $0 \leq \phi \leq 1$ indicates the fraction of one 
entire orbit completed. Eclipse of the x-ray source (i.e., superior 
conjunction of the neutron star or black hole) corresponds to $\phi = 0$. The
orbital period is 13.08 days.

\item The precession phase $0 \leq \psi \leq 1$ indicates the fraction of one 
entire jet precession cycle. $\psi=0$ is, by convention ({\it 2}), one of
the two instances when the beams of SS 433 are perpendicular to the line of
sight. The precessional period is 162.4 days. 

\item The Doppler factor $\delta = [\gamma(1-\beta \cos \theta)]^{-1}$ where
the velocity $v=\beta c$, $c$ is the speed of light, $\gamma =
(1-\beta^2)^{-1/2}$ is the Lorentz factor, and $\theta$ is the angle to the
line of sight. 

\item D. E. Harris and H. Krawczynski, {\it Astrophys. J.} {\bf 565},
244 (2002). 

\item M. Spada, G. Ghisellini, D. Lazzati, A. Celotti, {\it
Mon. Not. R. Astron. Soc.} {\bf 325}, 1559 (2001). 

\item G. V. Bicknell and D. B. Melrose, {\it Astrophys. J.} {\bf 262}, 511
(1982). 

\item The relationship $\Delta t=P(1-\delta)/2\delta$, where $P$ is the
precession period and $\delta$ is the fractional deviation from the mean
velocity $v_{m}$ of the ejected blobs, gives the time interval
$\Delta t$ over which a blob with velocity $(1+\delta)v_{m}$ reaches the blob launched with velocity
$(1-\delta)v_{m}$ at the same precessional phase one cycle earlier. Using
$P=162$ days and $\delta = 0.15$ we obtain $t\sim450$ days. 

\item Fig. 1 (top) is reproduced by permission of the
American Astronomical Society (Fig. 1 in ({\it 9})).   

\item We thank Dan Harris and Michiel van
der Klis for comments on a draft of this manuscript, and Wolfgang Brinkmann
and Mike Watson for earlier useful discussions. MM and RF are grateful to
Max-Planck-Institut f\"ur Astrophysik for their hospitality during the initial
analysis of these data.

\end{enumerate}
\end{quote}

\newpage

\begin{table}
\caption{Comparison of observed line centroid energies $E$ and
widths $\sigma$ with simulations based on symmetric jets integrated
over an entire precession cycle and a rest wavelength of 7.06 keV (Fe
XXV k$\beta$).}
\begin{tabular}{|ccccc|}
\hline
& $E_{\rm East}$ (keV) & $\sigma_{\rm East}$ (keV) & $E_{\rm West}$
(keV) & $\sigma_{\rm West}$ \\
\hline
Observed & $7.28^{+0.02}_{-0.23}$ & $0.7 \pm 0.2$ & $6.39^{+0.12}_{-0.15}$ & $0.2^{+0.2}_{-0.1}$ \\
\hline
Simulated & $7.40^{+0.1}_{-0.2}$ & $0.6 \pm 0.1$ & $6.40^{+0.1}_{-0.1}$ & $0.2^{+0.1}_{-0.2}$ \\
\hline
\end{tabular}
\end{table}

\newpage

\end{document}